# Accelerating the Kamada-Kawai algorithm for boundary detection in a mobile ad hoc network

SE-HANG CHEONG, University of Macau
YAIN-WHAR SI, University of Macau

Force-directed algorithms such as the Kamada-Kawai algorithm have shown promising results for solving the boundary detection problem in a mobile ad hoc network. However, the classical Kamada-Kawai algorithm does not scale well when it is used in networks with large numbers of nodes. It also produces poor results in non-convex networks. To address these problems, this paper proposes an improved version of the Kamada-Kawai algorithm. The proposed extension includes novel heuristics and algorithms that achieve a faster energy level reduction. Our experimental results show that the improved algorithm can significantly shorten the processing time and detect boundary nodes with an acceptable level of accuracy.

## 1. INTRODUCTION

Wireless sensor networks monitor and detect the physical environment with sensor nodes. A sensor node is a battery-powered device with special capabilities used to record or detect light, sound, humidity or the presence of certain objects. Sensors in close proximity to one another may also communicate through wireless communication (i.e., radio waves). Moreover, they can collect information and transmit the data back to central stations based on certain communication protocols. In general, wireless sensor networks consist of large numbers of sensors that are typically deployed within unattended environments. These sensors can operate autonomously. Due to the low costs associated with these sensors and because they are easy to maintain, wireless sensor networks are now widely used for reconnaissance during combat, security surveillance and disaster reporting and management. Recent developments in smart phone and mobile device technologies have also changed the traditional definition of a sensor node within an ad hoc network. Mobile devices can form ad hoc wireless sensor networks in a dynamic way and diverse applications can now be built to harness the processing power of these mobile devices.

Smart phones and mobile devices can be used to form ad hoc networks whenever telecommunication networks are damaged or become unavailable during disasters (e.g., during earthquakes and floods). In these situations, mobile devices within the ad hoc network can transmit emergency messages via nearby nodes to emergency service providers or rescue teams. Lifeline [1] is a prototype emergency ad hoc network that aims to enable the user to forward emergency messages based on a mobile ad hoc network protocol. The Lifeline App [1] is designed to instantly form an ad hoc emergency network using the Wi-Fi signals that originate from the mobile phones of disaster victims. It also uses signals from backup battery-powered wireless routers deployed by search and rescue teams. Such a system may be very useful in emergency situations. For example, a person trapped inside a confined area with no mobile network available could use the Lifeline App (previously downloaded to a mobile phone) to connect to an ad hoc network in cases where Lifeline is already installed on other devices. At this point, the victim can send messages to an emergency station via this wireless ad hoc network.

However, due to the nature of the ad hoc network, it can be difficult to establish a complete and stable network between all of the available mobile devices and routers. Therefore, a more effective emergency ad hoc network can be formed using several connected sub-networks. Some of these sub-networks can end up isolated and unable to communicate to the emergency stations or other networks. In such cases, emergency messages from mobile devices are not forwarded to the emergency stations on those networks. Furthermore, mobile devices have limited battery capacities. They cannot operate for more than a few days while processing/forwarding huge amounts of emergency messages to other nodes. Boundary nodes are among the more important elements in an emergency ad hoc network. They are the nodes most likely to connect to other nodes outside the ad hoc network. Therefore, their presence increases the chance of forwarding the messages to the nodes that belong to rescue personnel or emergency services. In addition, power management and message flow control are important issues that must be considered in any emergency ad hoc network. For instance, based on the information provided by the boundary nodes, various strategies may be used to extend battery life and prolong the ad hoc network's connection to other networks or nodes.

Boundary recognition and perimeter detection have both been extensively researched within the wireless sensor networks field. For example, recent reports have considered hole detection in ad hoc networks, topology estimations in ad hoc networks without physical locations, node position location and dynamic topology recovery. Classical force-directed algorithms cannot completely identify all of the boundary nodes in networks with complex topologies. However, these algorithms can still be useful for identifying boundary nodes with decent enough accuracy. The Kamada-Kawai with multiple node selection and decaying stiffness (KK-MS-DS) algorithm put forward in this paper focuses on the use of force-directed algorithms for boundary node detection in anchor-free networks. To reflect real situations, the proposed algorithm considers only edge length (signal strength) and node connection. In this paper, we introduce several heuristics, including multi-node selection, the use of signal strength for estimating node distance and decaying stiffness to achieve a fast convergence rate when detecting boundary nodes. In our experiments, we measured the sensitivity and specificity of the proposed algorithm by considering the different combination of the number of nodes and the average degree. Our main objective was to evaluate its effectiveness in various non-uniform and irregularly shaped networks for different node distributions.

In Section 2, we summarise the recent studies that have used geometric-, statistical- and topological-based methods to detect boundary nodes. In Section 3, we discuss the current well-known classical force-directed methods and evaluate their effectiveness in locating boundary nodes. In Section 4, we propose a heuristics algorithm (KK-MS-DS) for locating boundary nodes. In Section 5, we present the results of experiments conducted to evaluate the performance of the proposed algorithm. In Section 6, we conclude the paper and discuss future work.

## 2. RELATED WORK

Geometric-based methods require node position information. Fang et al. [2] proposed a distributed algorithm known as BOUNDHOLE for detecting the boundaries of inner holes. The algorithm can be applied to find an enclosed region within a network with no self-intersecting polygonal loop. It attempts to improve the quality of data transmitted into and across holes.

In addition, Fayed and Mouftah [3] proposed a heuristic algorithm known as local convex view (LCV) for locating boundary nodes. The LCV algorithm relies on a localised convex hull algorithm to identify boundaries. In this algorithm, every node calculates whether it is on the convex hull of its 1-hop neighbourhood. Fayed and Mouftah [4] also proposed an improved solution based on the alpha-shape principle. An alpha-shape is a geometric structure and a generalisation of the convex hull concept. It can capture the shape of a set of points along a plane [5]. Fayed and Mouftah [4] also proposed a localised boundary detection algorithm that relied on the hypothesis that some alpha-shapes are present in the view of the network. In their algorithm, a radius parameter α is defined and can be input by users to capture the alpha-shapes within a network. According to Fayed and Mouftah [4], alpha-shapes can also reveal inner and outer boundaries, which is beyond the ability of a general convex hull geometric structure.

Statistical-based methods for boundary detection work well when the density of the nodes in a network is sufficiently high. These methods require a certain number of samples and sufficient conditions for evaluation. Fekete et al. [6] described a method for identifying boundaries in a wireless network. In their approach, they assumed that nodes were located evenly over regions with a uniform distribution. They also assumed that nodes had a lower average degree along the boundaries and that interior nodes had higher average degrees. Moreover, Fekete et al. [7] proposed an algorithm for boundary recognition. Using this algorithm, they showed that restricted stress centrality was suitable for geometric sensor networks. They assumed that the nodes in a network were deployed uniformly and at random without location information. Their algorithm calculates the centrality index of the nodes within a network, where centrality indices are real value functions. In this case, the higher the values are, the more likely they are to be inner nodes. Conversely, the smaller the values, the more likely they are to be boundary nodes. The authors' study also analysed restricted stress centrality, which they calculated by measuring the total number of shortest paths passing through a node. Each node was then checked against the restricted street centrality above or below a threshold to determine whether they were inner or boundary nodes. However, statistical-based methods may not be suitable for ad hoc networks because these networks have varying numbers of nodes and densities and the issues of node distribution and edge connectivity can be quite complex. The nodes in a network may not be uniformly distributed, or the network may contain certain clusters.

Topological-based methods have also been used to determine boundaries without knowledge related to geometric locations and other information. However, these methods deal with complex combinatorial and computation problems in a distributed manner. Wang et al. [8] proposed an algorithm that detected the boundaries of holes using the shortest path tree. This algorithm builds the shortest path tree by flooding the network from an arbitrary root node upon initialisation. When there is no hole between the nodes within the shortest path tree, the shortest paths selected from the shortest path tree are more similar to straight lines. Otherwise, the shortest paths are curved. In this algorithm, distinct portions of similar paths that span the network are selected for detection [9].

Other studies have assumed that the communication network is a $d$-quasi unit disk graph model ($d$-QUDG). Here, $d$ is the longest Euclidean distance across which two nodes can communicate. If the Euclidean distance is longer than $d$, then the nodes cannot communicate. Saukh et al. [10] proposed an algorithm to find sub-graphs by locating chord-less cycles. The nodes used in their experiments were uniformly distributed at random with a unit disk graph. They merged the augmenting cycles around the interior nodes of their algorithm to form a boundary for the network. They proposed an algorithm according to the $\frac{\sqrt{2}}{2}$-QUDG model and made the communication distance $d$ easily adaptable to sparse networks.

Furthermore, Efrat et al. [11] proposed an approach to sensor localisation by using a force-direction algorithm when the length of the edge and angular information were already known. They used a multi-scale, dead-reckoning algorithm to extend the classical force-directed sensor localisation algorithms. Their study focused on the quality of sensor localisation for different varieties of non-convex shapes. The authors measured the performance and global quality of the output layout without noise. They also measured performance by simulating some noise on both edge lengths and angular information. Their approach focused on determining the layouts of network topology in non-convex shapes [11]. In contrast to their approach, the heuristics proposed in this paper focus on locating boundary nodes in sparse and non-uniform network topologies based on the Kamada-Kawai (KK) algorithm. In our method, we aim to achieve a rapid convergence rate in detecting boundary nodes rather than the quality of layout or the placement of inner nodes.

## 3. BOUNDRY DETECTION IN AN AD HOC NETWORK WITHOUT LOCATION INFORMATION USING A FORCE-DIRECTED APPROACH

In general, each node within a mobile ad hoc network can be specified by their (partial) geographic location. A mobile ad hoc network can be set up with custom-made hardware such as GPS, or enhanced antennas with angle information can be embedded within the nodes. However, building a large mobile ad hoc network that embeds positioning hardware within the nodes would be impractical. It would be neither cost effective nor energy efficient to do so, especially in temporary networks and indoor environments [12]. When the nodes within mobile ad hoc networks do not have any geographic locations, other information such as the network topology, signal strength of the connections and information about neighbouring nodes are often the only kind of data available to describe the entire network. Boundary detection often relies on information obtained by broadcasting messages across the network and reasoning from reply messages. However, there are a number of challenges associated with such approaches. For instance, a node within the mobile ad hoc network may not be able to handle large amounts of traffic due to its limited processing capability. Therefore, it could become a bottleneck for the entire network when it is chosen to broadcast or forward the network traffic.

The network topology of a mobile ad hoc network can take many forms. For example, Figure 1(a) and Figure 1(b) have the same network topology, although they can be drawn in different ways. In Figure 1(b), the network is drawn without a crossing edge. However, there are two crossing edges in Figure 1(a). Some algorithms are capable of drawing graphs based on a given topology. For instance, force-directed graph-drawing algorithms are widely used in data visualisation. These algorithms have been studied in the contexts of both sensor and ad hoc networks. They are designed to position the nodes and edges so that there are as few crossing edges as possible based on the given topology of the edges and nodes.

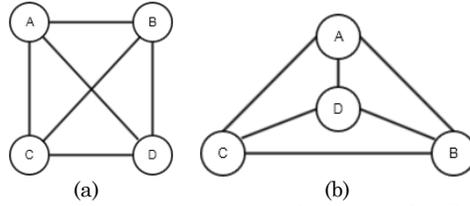

(a) (b)
Figure 1(a) Representation A of a network topology and (b) Representation B of a network topology.

### 3.1 Evaluation of the current force-directed approaches

In the following sections, we review three well-known force-directed algorithms and analyse their performance in terms of their ability to locate boundary nodes. The key objective of the graph-drawing methods in force-directed algorithms is to distribute the nodes and edges uniformly and symmetrically whenever possible. In classic force-directed methods, the new position of a node within a network (graph) is determined iteratively. During this process, the algorithm calculates local and/or global forces of attraction and repulsion iteratively. The location of the nodes is then adjusted based on the forces between them and their related neighbours. Nodes that are connected via an edge attract one another; otherwise, they repel one another. The basic idea is that adjacent nodes must be placed close together (but not too close) and clusters must be separated from other clusters to avoid clutter. In our approach, we attempted to locate the boundary nodes of an ad hoc network. We considered a network to be connected when every node could reach another node via edges. We did not consider isolated nodes (i.e., nodes that had no edge connection to other nodes) in our approach.

In the following sections, we evaluate the current force-directed algorithms for addressing the boundary detection problem. In these experiments, we evaluated their true positive and false negative rates for locating boundary nodes. The main objective of this evaluation was to determine how accurate the current force-directed algorithms were when applied to the boundary detection process. In our evaluation, we executed these algorithms until the termination conditions were met. To achieve fair evaluations for these algorithms, a randomised series of network topologies were used for testing. Finally, we compared the boundary nodes from the initial topology and boundary nodes returned by the algorithms. These experiments were performed with a computer containing an Intel Pentium T2390 processor and 4 GB of memory. $n$ is the number of nodes in the network, $\delta$ is the distribution ratio of nodes, $\gamma$ is the edge distribution ratio and $\gamma_b$ is the probability of the generation of an edge $b$. The topology generation procedure for the evaluation is written as follows.

1. We set the initial parameters such as the node count ($n$), node distribution ratio ($\delta$), edges radio ($\gamma$) and edge connectivity ($\gamma_b$).
2. The node count ($n$) was chosen from the range of [10, 1,000].
3. We set the node distribution ratio ($\delta$) to $1.7 \times \frac{1}{\sqrt{n}}$, the edge radio ($\gamma$) to $0.7 \times \frac{1}{\sqrt{n}}$ and the edge connectivity ($\gamma_b$) to 0.7.

We generated 991 network topologies with a node count ($n$) from 10 to 1,000. The average degree of the 991 network topologies was 7.236054. The maximum and minimum degrees of the 991 network topologies were 8 and 3, respectively.

**ALGORITHM 1.**   Pseudo code of network generation

*Initialise parameters* $n, \delta, \gamma, \gamma_b$;

```
for i = 10 to 1,000 do {
    g = network_generator(n, δ, γ, γ_b);
    save g;
}
```

Algorithm 1 Pseudo code of network generation

### 3.2 Kamada-Kawai

The KK algorithm [13] is based on Eades' spring-embedder model [14]. It keeps edge crossing to a minimum and distributes the nodes and edges uniformly [15]. However, the KK algorithm is much slower than the other force-directed algorithms. Furthermore, it cannot be applied to a network containing more than 500 nodes [16]. In the KK algorithm, nodes are placed so that their visual distance within the drawing is proportional to their theoretical graphed distance. As this goal cannot always be achieved for arbitrary network topologies, the key idea behind the algorithm is to use a spring model in such a way that minimises the energy function of the network topology. The energy function $E$ is:

$$E = \sum_{i=1}^{n-1} \sum_{j=i+1}^{n} \frac{1}{2} k_{i,j} (|p_i - p_j| - l_{i,j})^2 \quad (1)$$

where $k_{i,j}$ is the stiffness of a spring between nodes $i$ and $j$, $l_{i,j}$ is the ideal distance of a spring between nodes $i$ and $j$ and $p_i$ and $p_j$ are the visual positions of nodes $i$ and $j$, respectively. That is, the KK algorithm finds a visual position for each pair of nodes $i$ and $j$, and their Euclidean distance is proportional to $l_{i,j}$. Here, the KK algorithm defines a diameter matrix that stores all of the theoretical graphed distances ($d_{i,j}$) between all of the nodes. $d_{i,j}$ represents the hop count between nodes $i$ and $j$. $d_{i,j}$ records the shortest hop count between nodes $i$ and $j$ from all of the possible paths. The ideal distance of a spring ($l_{i,j}$) between nodes $i$ and $j$ can then be defined as follows:

$$l_{i,j} = \frac{L_0}{\max_{i<j} d_{i,j}} \times d_{i,j} \quad (2)$$

where $L_0$ is the side length of the drawing frame and $\max_{i<j} d_{i,j}$ is the diameter of the network topology. Moreover, the stiffness of a spring between nodes $i$ and $j$ is calculated as:

$$k_{i,j} = \frac{K}{d_{i,j}^2} \quad (3)$$

where K is a scale constant and $d_{i,j}$ represents the theoretical graphed distances of nodes $i$ and $j$.

The KK algorithm then seeks a visual position for every node $v$ in the network topology and tries to decrease the energy function in the whole network. That is, the KK algorithm calculates the partial derivatives for all of the nodes in the network topology in terms of every $x_v$ and $y_v$ that are zero (i.e., $\frac{\partial E}{\partial x_v} = 0$ and $\frac{\partial E}{\partial y_v} = 0$, for $1 \leq v < n$).

However, solving all of these non-linear equations simultaneously is unfeasible because they are dependent on one another. Therefore, an iterative approach can be used to solve the equation based on the Newton-Raphson method. At each iteration, the algorithm chooses a node $m$ that has the largest maximum change ($\Delta_m$). In other words, the node $m$ is moved to the new position, where it can reach a lower level of $\Delta_m$ than before. Meanwhile, the other nodes remain fixed. The maximum change ($\Delta_m$) is calculated as follows:

$$\Delta_m = \sqrt{\left(\frac{\partial E}{\partial x_m}\right)^2 + \left(\frac{\partial E}{\partial y_m}\right)^2} \qquad (4)$$

The pseudo code for the KK algorithm [13] is given in Algorithm 2.

---

**ALGORITHM 2.**   Pseudo code of Kamada-Kawai algorithm

*Input*: network topology $G = (V, E)$
*Output*: a visual drawing of $G$
compute theoretic graphed distance $d_{i,j}$ for $1 \le i \ne j \le n$;
compute ideal distance $l_{i,j}$ for $1 \le i \ne j \le n$;
compute stiffness $k_{i,j}$ for $1 \le i \ne j \le n$;
initialise position for node $1, 2, \ldots n$;
**while** $max_i \Delta_i > \varepsilon$ **do** {
    let $node_m$ be the node satisfying $\Delta_m = max_i \Delta_i$;
    **while** $\Delta_m > \varepsilon$ **do** {
        compute $\delta x$ and $\delta y$ for $node_m$;
        $x_m = x_m + \delta x$; /* update the x position of $node_m$ */
        $y_m = y_m + \delta y$; /* update the y position of $node_m$ */
    }
}

Algorithm 2 Pseudo code of Kamada-Kawai algorithm

---

In this section, we evaluate the performance of the KK algorithm given in Algorithm 2 in terms of boundary detection. In our evaluation, we set $K$ to 1. The algorithm was allowed to execute for up to 5 minutes or until the value of energy function $E$ was less than 1.0.

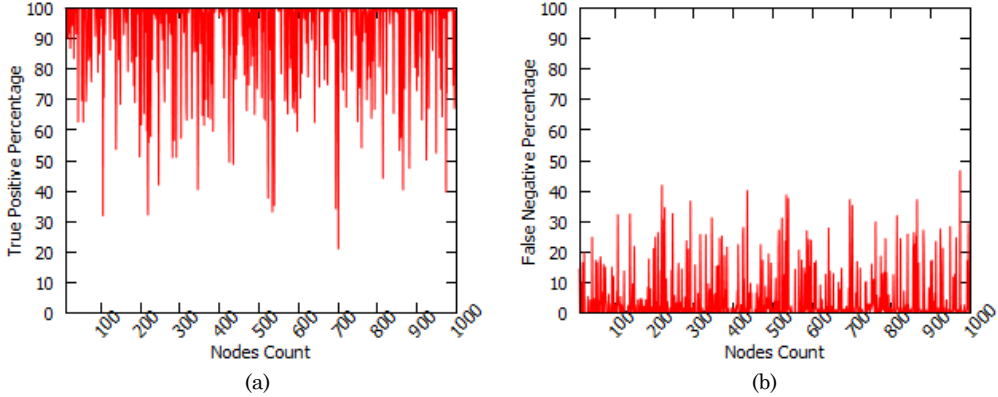

(a)          (b)
Figure 2(a) Kamada-Kawai (true positive rate) and (b) Kamada-Kawai (false negative rate).

Figure 2(a) shows the true positive rate of the KK algorithm. From the results, we can see that most of the test cases had a 60% or higher true positive rate. We can also observe that the true positive rate for the KK algorithm did not decrease significantly when the number of nodes increased. Figure 2(b) shows the false negative rate using the KK algorithm. The results here also reveal that for most of the network topologies, the false negative rate was lower than 30% and did not depend on the number of nodes or average degree. That is, both the true and false positive rates remained stable regardless of the different parameters.

### 3.3 Fruchterman Reingold

The Fruchterman Reingold [17] algorithm is based on Eades' spring-embedder model [14]. It distributes nodes evenly while minimising edge crossings. It also maintains uniform edge lengths. In contrast to the KK algorithm, it uses two forces (attraction and repulsive forces) to update nodes rather than using an energy function with a theoretical graphed distance.

First, the attraction force ($f_a$) and repulsive force ($f_r$) are defined as follows:

$$f_a(d) = \frac{d^2}{k} \quad (5)$$

$$f_r(d) = -\frac{k^2}{d} \quad (6)$$

where $d$ is the distance between two nodes and $k$ is a constant of ideal pairwise distance. The constant of ideal pairwise distance ($k$) of attraction force ($f_a$) and repulsive force ($f_r$) are $a \times \sqrt{\frac{W \times H}{n}}$ and $r \times \sqrt{\frac{W \times H}{n}}$, respectively. Here, $W$ is the width of the drawing frame, $H$ is the height of the drawing frame, $n$ is the total number of nodes in the network topology, $a$ is a constant for the attraction multiplier and $r$ is a constant for the repulsive multiplier.

The Fruchterman Reingold algorithm is executed iteratively and all of the nodes are moved simultaneously after the forces are calculated for each iteration. The algorithm adds an attribute 'displacement' to control for the position offset of the nodes. At the beginning of an iteration, the Fruchterman Reingold algorithm calculates the initial value of the displacement for all of the nodes with the use of repulsive force ($f_r$). The algorithm also uses attraction force ($f_a$) to iteratively update the visual position of the nodes on every edge. Finally, it updates the position offset of the nodes using the displacement value.

The displacement scale $s$ is used as the termination condition of the Fruchterman Reingold algorithm. When the displacement scale ($s$) is less than the threshold $\varepsilon$, the algorithm can terminate. Upon initialisation of the algorithm, the displacement scale ($s$) is assigned to $\frac{W}{10}$. This value is updated at each iteration according to the iteration count and maximum iteration input by users. The pseudo code for the Fruchterman Reingold algorithm [17] is given in Algorithm 3.

**ALGORITHM 3.** Pseudo code of Fruchterman Reingold algorithm

**Input**: $network\ topology\ G = (V, E)$
**Output**: $a\ visual\ drawing\ of\ G$
$initialise\ the\ iteration\ count\ it = 0;$
$s = \frac{W}{10};$
**while** $s \geq \varepsilon$ **do** {
　　$/*\ calculate\ the\ initial\ value\ of\ displacement\ */$
　　**foreach** $u \in V$ **do** {
　　　　$u_{displacementX} = 0;$
　　　　$u_{displacementY} = 0;$
　　　　**foreach** $v \in V, u \neq v$ **do** {
　　　　　　$\Delta = \sqrt{(u_x - v_x)^2 + (u_y - v_y)^2};$
　　　　　　$u_{displacementX} \mathrel{+}= \frac{u_x - v_x}{\Delta} \times f_r(\Delta);$

$$u_{displacementY} \mathrel{+}= \frac{u_y - v_y}{\Delta} \times f_r(\Delta);$$
}
}

/∗ *update the value of displacement* ∗/
**foreach** $e \in E$ **do** {
  Let $u, v$ be the end nodes of $e$;
  $$\Delta = \sqrt{(e.u_x - e.v_x)^2 + (e.u_y - e.v_y)^2};$$
  $$e.u_{displacementX} \mathrel{-}= \frac{u_x - v_x}{\Delta} \times f_a(\Delta);$$
  $$e.u_{displacementY} \mathrel{-}= \frac{u_y - v_y}{\Delta} \times f_a(\Delta);$$
  $$e.v_{displacementX} \mathrel{+}= \frac{u_x - v_x}{\Delta} \times f_a(\Delta);$$
  $$e.v_{displacementY} \mathrel{+}= \frac{u_y - v_y}{\Delta} \times f_a(\Delta);$$
}

/∗ *update the position of nodes* ∗/
**foreach** $v \in V$ **do** {
  $$\Delta = \sqrt{v_{displacementX}^2 + v_{displacementY}^2};$$
  $$v_x = v_x + \frac{v_{displacementX}}{\Delta} \times min(v_{displacementX}, s);$$
  $$v_y = v_y + \frac{v_{displacementY}}{\Delta} \times min(v_{displacementY}, s);$$
  $$v_x = min\left(\frac{W}{2}, max\left(-\frac{W}{2}, v_x\right)\right);$$
  $$v_y = min\left(\frac{H}{2}, max\left(-\frac{H}{2}, v_y\right)\right);$$
}

/∗ *update displacement scale* ∗/
$it = it + 1;$
$$s = s * \left(1 - \frac{it}{\max iteration}\right);$$
}

Algorithm 3 Pseudo code of Fruchterman Reingold algorithm

Similar to the KK algorithm, we evaluated the application of the Fruchterman Reingold algorithm to the boundary detection problem. We set the attraction multiplier ($a$) to 0.75 and the repulsion multiplier ($r$) to 0.75 in the Fruchterman Reingold algorithm. The algorithm was allowed to execute for up to 5 minutes or until the value of the displacement scale ($s$) was less than or equal to $1 \times 10^{-6}$. These experiments were conducted with the same network topologies used when testing the KK algorithm.

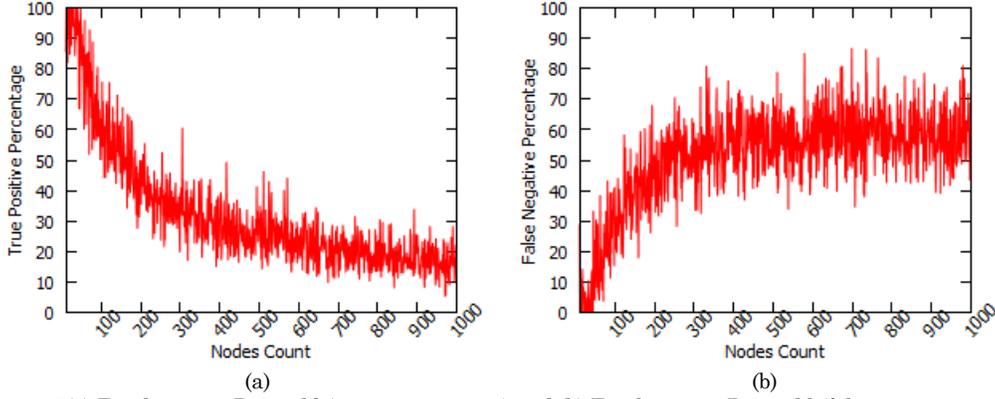

(a) (b)
Figure 3(a) Fruchterman Reingold (true positive rate) and (b) Fruchterman Reingold (false negative rate).

Figure 3(a) illustrates the true positive rate for detecting boundary nodes using the Fruchterman Reingold algorithm. The true positive rate decreased when the number of nodes in the test case increased. However, when the number of nodes was less than 150, the true positive rate was 60% or higher. Meanwhile, when the number of nodes was greater than 400, the true positive rate decreased to less than 50%. Figure 3(b) illustrates the false negative rate using the Fruchterman Reingold algorithm. The results are similar to the complement of the true positive rate. When the number of nodes was less than 150, the percentage of the false negative was relatively small and lower than 30. However, when the number of nodes was larger than 250, the false positive rate increased to more than 50% for most of the cases.

### 3.4 Davidson Harel

The Davidson Harel algorithm [18] distributes nodes evenly while minimising edge crossings. It also prevents nodes from getting too close to non-adjacent edges. The Davidson Harel algorithm applies a simulated annealing process to graph the drawing. In doing so, it ensures that the search process is not constrained at a local minimum. This process is based on a similar analogy to the physical annealing process in which liquids are cooled into a crystalline form.

The energy value $E$ of the initial network topology for the simulated annealing process was calculated upon initialisation. Here, we selected attraction force $d_a$ and repulsive force $d_r$ as the energy functions of the Davidson Harel algorithm.

$$d_a(y) = y^2 \qquad (7)$$

$$d_r(y) = \frac{1}{(y+1)^2} \qquad (8)$$

where $y$ is the Euclidean distance between two nodes. We iterated all of the pairs of nodes in the network topology and calculated the attraction force $d_a$ and repulsive force $d_r$ for each. An initial energy value ($E$), which is the sum of all of the attraction forces ($d_a$) and repulsive forces ($d_r$), is calculated as follows:

$$E = \sum_{i=1}^{n-1}\sum_{j=i+1}^{n} d_a\left(\sqrt{(i_x - j_x)^2 + (i_y - j_y)^2}\right) + d_r\left(\sqrt{(i_x - j_x)^2 + (i_y - j_y)^2}\right) \qquad (9)$$

The Davidson Harel algorithm executes iteratively. A node $i$ is randomly selected from the network topology at the beginning of each iteration. The algorithm then creates a temporary node $j$ with a new position assigned based on the position of node $i$. A new energy value $E'$ is then calculated based on the position of node $j$ and the position of nodes within the network topology.

$$E' = \sum_{v,i \in V, j \notin V, v \neq i} d_a\left(\sqrt{(v_x - j_x)^2 + (v_y - j_y)^2}\right) + d_r\left(\sqrt{(v_x - j_x)^2 + (v_y - j_y)^2}\right) \quad (10)$$

In the Davidson Harel algorithm, the simulated annealing obeys Boltzmann distribution rules when the liquid is cooled slowly (i.e., it reaches thermal equilibrium at every temperature) [19]. Once the new energy ($E'$) is calculated on an iteration, either $E'$ or $E$ is selected for the next iteration. That is, if $E' - E \leq 0$, then the new energy ($E'$) replaces the old energy ($E$) on the next iteration because the new energy brings the system to a lower energy. If $E' - E > 0$, then the new energy brings the system to a higher level of energy. However, the new energy is not rejected directly, but rather accepted with the following probability:

$$p = e^{-\frac{(E'-E)}{k \times T}} \quad (11)$$

where $T$ is the temperate variable and $k$ is the Boltzmann constant. In our implementation, to accept a higher energy ($E'$) with a probability, we generated a random number $\varphi$ between 0 and 1. If the probability ($p$) was less than the random number ($\varphi$), we then accepted the new energy ($E'$); otherwise, we kept the old energy ($E$) on the next iteration.

The Davidson Harel algorithm terminates if the temperature ($T$) is less than a threshold $\varepsilon$. The pseudo code for the Davidson Harel algorithm is given in Algorithm 4.

**ALGORITHM 4.**   Pseudo code of Davidson Harel algorithm

*Input*: *network topology* $G = (V, E)$
*Output*: *a visual drawing of* $G$
*initialise the temperature* $t$;
*initialise the iteration of a temperature* $itmax = 20 \times number\ of\ nodes\ in\ G$;
*initialise the cooling constant of temperature* $t_c$;
*initialise the radius of disk* $r$;
*initialise the shrinking constant of radius of disk* $r_c$;

*compute the energy value* $E$;
**while** $t \geq \varepsilon$ **do** {
    **for** $i = 1$ **to** $itmax$ **do** {
        /* *initialise a temporary node* */
        *let* $node_i \in V$ *be the node selected randomly*;
        *let* $node_j \notin V$ *be the temporatory node*;
        $angle = random(2 \times \pi)$;
        $node_j.x = node_i.x + \cos(angle) \times r$;
        $node_j.y = node_i.y + \sin(angle) \times r$;

        /* *compare the new energy value E' and old energy value E* */
        *compute the energy value* $E'$;
        **if** $E' \leq E$ **then** {
            $node_i.x = node_j.x$;
            $node_i.y = node_j.y$;
            $E = E'$;
        } **else** {
            *compute* $p$;
            $\varphi = random(0,1)$;

```
            if p < φ then {
                node_i.x = node_j.x;
                node_i.y = node_j.y;
                E = E';
            }
        }
    }
    t = t × t_c;
    r = r × r_c;
}
```

Algorithm 4 Pseudo code of Davidson Harel algorithm

We evaluated the performance of the Davidson Harel algorithm as applied to boundary detection by setting the disk radius to 100, the cooling factor to 0.75, the shrink factor to 0.8 and the temperature to 1,000. The algorithm executed for up to 5 minutes or until the value of temperature $t$ was less than or equal to 1.0. We evaluated the algorithm based on the same network topologies used in the testing of the KK and Fruchterman Reingold algorithms.

(a)

(b)

Figure 4(a) Davidson Harel (true positive rate) and (b) Davidson Harel (false negative rate).

Figure 4(a) illustrates the true positive rate using the Davidson Harel algorithm. The experimental results reveal that the true positive rate was stable at around 20% when the number of nodes was greater than 250. When the number of nodes was less than 100, a 40% true positive rate was achieved. Figure 4(b) illustrates the false negative rate. Figure 4(b) indicates that the trend was similar to the results obtained for the true positive rate. Approximately 30-40% of the nodes were incorrectly identified as boundary nodes when the number of nodes was greater than 250.

### 3.5 Summary of Evaluation Results

Figure 5 illustrates the overall results for our evaluation of the three force-directed algorithms.

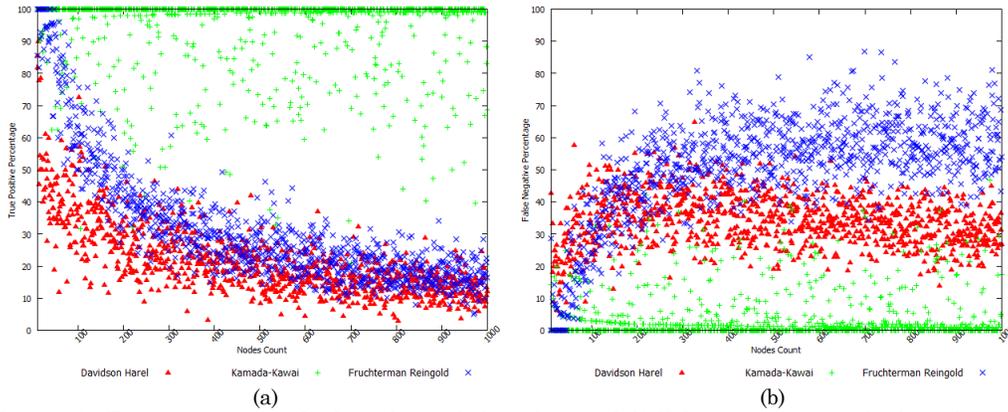

Figure 5(a) True positive rates for force-directed algorithms and (b) False negative rates for force-directed algorithms

According to the evaluation results, the average true positive rates for the KK, Fruchterman Reingold and Davidson Harel algorithms were 93.95226%, 32.30388% and 21.93083%, respectively. The average false negative rates were 3.82441%, 50.88479% and 34.04234%, respectively. In addition, the Fruchterman Reingold algorithm had a higher true positive rate when the number of nodes was small. However, it had a high false negative rate when the number of nodes was large. Compared with the other two algorithms, the Fruchterman Reingold algorithm had the highest false negative rate. Furthermore, although the true positive rate of the Davidson Harel algorithm was low, it had stable true positive and false negative rates. It was also the most stable algorithm out of the three considered. Moreover, the KK algorithm had the highest true positive percentages and the lowest false negative rates. It also showed varying results for some of the input topologies.

In conclusion, the KK algorithm demonstrated the highest level of precision out of the three algorithms tested for boundary detection. However, one of the disadvantages associated with the KK algorithm is that it can be very slow for large networks and its overall results may be considered unstable. To alleviate these problems, we propose a new and faster KK algorithm for boundary detection based on multi-node selection and the incorporation of a decaying stiffness calculation.

## 4. USING KAMADA-KAWAI FOR BOUNDARY DETECTION IN MOBILE AD HOC NETWORKS

In the following sections, we introduce a number of improvements to the KK algorithm for addressing the boundary detection problem in mobile ad hoc networks. These improvements include an adoption of signal strength and stiffness, a batch updating of nodes and the use of decaying stiffness to accelerate the convergence rate.

### 4.1 Kamada-Kawai with signal strength (KK-SS)

One of the characteristics of an ad hoc network is that every node within the network has its own specific transmission range. The nodes that are connected to one another must be located within their specified ranges. Based on this property, we can consider the edge lengths of nodes as being proportional to their signal strengths. In the KK algorithm, there is a diameter matrix that stores all of the theoretical graphed distances $(d_{ij})$, as described in Section 3.2. Based on this assumption, we can extend the KK algorithm by substituting the values in the diameter matrix with the estimated distances calculated by the signal strength. To achieve this objective, we use free-space path loss (FSPL), which calculates the estimated distance based on signal strength. FSPL is a term used in the telecommunications field to denote the loss in signal strength of an electromagnetic wave as a result of a line-of-sight path. The path usually travels through a free space (air) with the assumption that there are no obstacles nearby to cause reflection or diffraction. In this paper, we adopt the FSPL assumption to estimate the distance between nodes. However, we assume an ideal situation. In practice, the estimation of distance can be affected if the nodes are placed indoors or the signals reflect off walls. The estimated distance ($d$) in meters can be calculated as follows:

$$d = 10^{\frac{27.55 - (20 \times \log_{10}(f)^{-s})}{20}} \qquad (12)$$

where $f$ is the signal strength and $s$ is the signal frequency in Mhz.

### 4.2 Kamada-Kawai with multi-node selection (KK-MS)

In this section, we introduce a heuristic for accelerating the KK algorithm. Our aim is to optimise the energy function of KK by adopting the characteristics of mobile ad hoc networks. Recall that the KK algorithm selects and updates the node with the maximum change ($\Delta_m$) as described in equation (4) in one iteration. However, the experiments we conducted using the KK algorithm with a constant execution time revealed that the larger the topology, the lower the true positive rate. Therefore, we conclude that the rate of the KK algorithm's node updating is too slow, especially for large topologies. As updating is done one node at a time in the KK algorithm, more iterations are needed when the topology is large, and thus it takes longer to execute.

Against this background, we introduce a heuristic known as multi-node selection that updates a group of $k$ nodes at every iteration in the KK algorithm. In the initialisation step, we calculate the maximum change ($\Delta_m$) for every node and insert the top-$k$ nodes into an ordered queue. Next, we pop up the first $max\ k$ nodes from the queue and update them. We then recalculate their maximum change ($\Delta_m$) values and put them back into the queue for the next round. We update the maximum change ($\Delta_m$) for all of the nodes when $\sqrt{n}$ nodes have been selected (where $n$ is the total number of nodes in the network). To ensure that the nodes from different parts of the network have been properly selected, the algorithm removes a node from the list if it is within three hops of the selected node. The pseudo code of the KK with multi-node selection (KK-MS) algorithm is given in Algorithm 5.

---

**ALGORITHM 5.** Pseudo code of Kamada-Kawai with multi-node selection

---

**Input**: $network\ topology\ G = (V, E), percentage\ of\ node\ for\ selection\ k$
**Output**: $a\ boundary\ ready\ network\ topology\ of\ G$
$compute\ theoretic\ graphed\ distance\ d_{i,j}\ for\ 1 \leq i \neq j \leq n;$
$compute\ ideal\ distance\ l_{i,j}\ for\ 1 \leq i \neq j \leq n;$
$compute\ stiffness\ k_{i,j}\ for\ 1 \leq i \neq j \leq n;$
$initialise\ position\ for\ node\ 1, 2, \dots n;$
$initialise\ an\ ordered\ queue\ Q = \{nodes\ with\ top_k\ \Delta_v\ for\ v \in V\ \};$
**while** $max_i \Delta_i > \varepsilon$ **do** {
    **if** $\sqrt{n}$ nodes have been selected **then** {
        $clear\ Q;$
        **foreach** $v \in V$ **do** {
            **if** $\Delta_v\ is\ in\ top_k\ maximum\ change$ **and** $hopcount(Q, v) > 3$ **then** {
                $push\ v\ into\ Q;$
            }

```
        }
    }
    foreach v ∈ Q do {
        while (Δ_v > ε) do {
            compute δx and δy for node_v;
            x_v = x_v + δx; /* update the x position of node_v */
            y_v = y_v + δy; /* update the y position of node_v */
        }
    }
}
```

Algorithm 5 Pseudo code for Kamada-Kawai with multi-node selection

### 4.3 Kamada-Kawai with multi-node selection and decaying stiffness (KK-MS-DS)

Recall that our aim is to achieve fast convergence of the KK algorithm on boundary node detection. Therefore, fitting to a specified drawing frame and achieving a nice output layout are not among our major concerns. When accelerating the KK algorithm, the network topology must be built into a coarse enough layout and the nodes must be updated as rapidly as possible upon initialisation. It is also crucial to determine a stopping mechanism to prevent the algorithm from spending too much time on nodes that are already placed near or at their final positions in the graph.

Although an increase in the number of iterations can accelerate the movement of the nodes, it can also significantly increase the execution time. Updating the position of a selected node many times in one iteration may result in little improvement as the node energy decreases iteratively in the KK algorithm and the node can still be shifted at a slower pace.

The KK algorithm uses stiffness to control the movement of the node as described in equation (3). Reducing the stiffness value reduces the level of energy for a pair of nodes, as described in equation (1). Therefore, we conducted an experiment to evaluate the effect of stiffness on different drawing frames with varying sizes while keeping the stiffness coefficient at 1.0. We then compared the results with those of the original KK algorithm. In the experiment, we used the same settings used in Section 3.1. However, we set the termination condition of the algorithm to 60% of the sensitivity or an energy level less than or equal to 1.0. The algorithm terminated once one of these conditions is met.

Figure 6 presents our experimental results. Although modified stiffness may achieve a shorter execution time, it may cause a higher false positive rate on some network topologies. Furthermore, the change in stiffness can affect the execution time and cannot guarantee a nice placement of inner nodes. Moreover, there are other disadvantages of using constant stiffness in some of the network topologies. Setting the stiffness to 1.0 also produced poor results on non-convex and large networks, especially for network topologies with inner holes. The false positive rate for these network topologies was very high, and the nodes within the network were stacked together during the iteration.

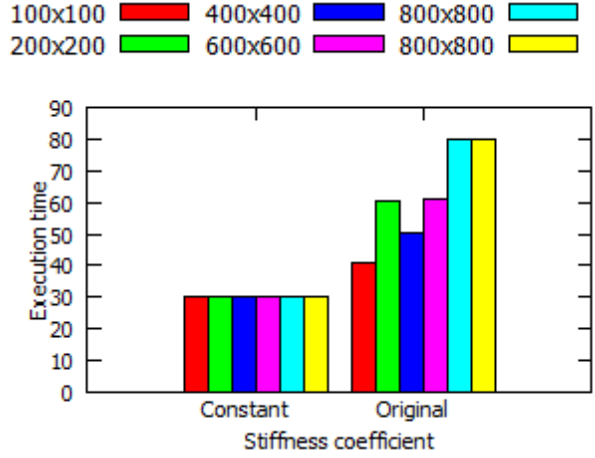

Figure 6 Constant stiffness coefficient vs. original stiffness coefficient

The KK algorithm can quickly achieve a stable state if the position update on the nodes is sufficient. A certain scale of network topologies must take certain iterations to become stable. A stable status means that the sensitivity and specificity do not improve for a predefined number of iterations. Although further refinement may contribute only a little (or nothing at all) to sensitivity and specificity, it requires consuming a disproportionate amount of execution time.

According to the preceding analysis, we decided against the use of a constant value for stiffness. Instead, we introduce a decaying stiffness ($m$) for every node that replaces the node stiffness in the KK algorithm. Decaying stiffness ($m$) is a real number with a range from 0 to 1. In general, the higher the value of decaying stiffness ($m$) a node has, the further the distance the node can be moved. The main objective of the proposed approach is to achieve fast convergence for boundary node detection by pushing the nodes in the outer boundary away from the inner nodes as much as possible, regardless of the placement of the inner nodes and the quality of the layout. The proposed approach includes the following four steps.

1. **Select an area from the network topology upon initialisation**

Upon initialisation, we select a node with the highest average degree to be the preferred starting point. Next, we collect all of its two-hop nodes and construct an initial starting area. We then execute the KK-MS algorithm on the starting area with 5% of the nodes selected simultaneously. Furthermore, we calculate the sensitivity and specificity of the starting area after a predefined iteration. In our experiment, we used 100 iterations.

2. **Assign decaying stiffness to existing nodes from the starting area**

We use an exponential decay function to update the decaying stiffness ($m$) of the nodes from the starting area. The value of decaying stiffness ($m$) of the nodes decreases along with the execution time spent on the starting area. The value of the updated decaying stiffness ($m'$) is calculated as

$$m´ = m - zp^t \qquad (13)$$

where $z$ is the normalised value (values range from 0 to 1) of the maximum change ($\Delta_m$), which is calculated by equation (4). $p$ is the decay rate (we set $p$ to 0.05 in our experiments), and $t$ is the number of times the node has been selected for updating. The higher the decay rate, the more likely that the node will be selected for the next iteration.

### 3. Adding outside nodes into the starting area

Every time the ratio of the stable status ($r$) of the starting area does not improve for a predefined number of iterations, the starting area is expanded by adding new nodes from the neighbouring area into the starting area. In our approach, we use a heuristics-based estimation to evaluate the stable status of the starting area. A stable status implies that the starting area has been formed up to a coarse layout. This means that the layout cannot 'expand' any further using the nodes in the starting area. At this time, the neighbouring nodes surrounding the starting area are added into the starting area. In our experiment, we check the stable status at every 100 iterations based on equation (14).

$$ME = \frac{1}{l}\sum_{i=1}^{l} \hat{L}_i - L_i \tag{14}$$

where $l$ is the total number of edges in the network topology, $\hat{L}_i$ is the edge length on the intermediate iteration and $L_i$ is the edge length on the input network topology (an estimated distance calculated by equation (12)). The ratio of the stable status ($r$) of the starting area is given by:

$$r = \frac{\frac{1}{l}\sum_{i=1}^{l}|\hat{L}_i - L_i|}{\sigma}, \sigma = \sqrt{\sum_{i=1}^{l}(\hat{L}_i - ME)^2} \tag{15}$$

where σ is the deviation of the difference in the estimated distance on the input network topology and an intermediate iteration. Therefore, when the ratio of the stable status ($r$) is lower than a threshold $\varepsilon$, the current starting area is marked as stable and rebuilt.

The stiffness of the newly added nodes is replaced by the decaying stiffness ($m$) with a value of 1.0. The larger the value of the decaying stiffness, the higher the chance the node has of being selected for a position update. Furthermore, the higher the value of the decaying stiffness of a selected node, the larger its degree of movement.

The other nodes inside the starting area are set to a small value of decaying stiffness. In our experiment, we use 0.1 on these nodes. We assign this small value of decaying stiffness because these nodes have been moved (selected) in previous iterations and the sensitivity and specificity of the starting area cannot be improved any further. As mentioned previously, when the algorithm does not execute enough iterations for the nodes within the network topology, the layout of the network may be unable to 'expand' in an outward direction effectively. Therefore, we select a starting area in the input network topology and expand it as quickly as possible to cover the entire network.

### 4. Fine-tuning the network topology

When the starting area covers the entire network and becomes stable, this implies that the nodes in the whole network cannot be further adjusted substantially. Therefore, a final step in the fine-tuning procedure is used to improve the final position of the nodes. In this procedure, we bring the stiffness of all of the nodes in the network back to $k_{i,j}$ which can be calculated using equation (3). We then recalculate the ratio of the stable status ($r$) of the entire network using equation (15). The fine-tuning step terminates if the ratio of the stable status ($r$) is less than the threshold $\varepsilon$ or it remains unchanged up to certain period. Once the fine-tuning is terminated, the network is ready for the boundary node detection process.

The pseudo code for the KK-MS-DS algorithm (from steps 1 to 4) is given in Algorithm 6.

**ALGORITHM 6.** Pseudo code of Kamada-Kawai with multi-node selection and decaying stiffness

*Input*: network topology $G = (V, E)$
*Output*: a boundary ready network topology of $G$
initialise the staring area $WT = (V_{WT}, E_{WT})$ that $WT \subseteq T$;
initialise the iteration count $it = 0$;
initialise the iteration to calcualte $r$ as $tt = 100$;
initialise the starting node $s$ which has a maximum average degree in $G$;

```
/* step 1 */
add node s into V_WT;
foreach v ∈ V do {
    if hopcount(V_WT, v) ≤ 2 then {
        add v into V_WT;
        v_stiffness = v_decaying stiffness;
    }
}

/* step 2 and step 3 */
while WT ≠ T do {
    it = it + 1;
    compute r for the WT every tt iteration;

    /* the starting area is stable */
    if r < ε then {
        foreach v ∉ V_WT, v ∈ V do {
            if hopcount(V_WT, v) ≤ 2 then {
                add v into V_WT;
                v_stiffness = largest stiffness from its neighbours;
            } else {
                v_stiffness = K / d_{i,j}^2;
            }
        }
        foreach u, v ∈ V_WT, u ≠ v do {
            update L̂_i for node v and v;
        }

    /* the starting area is not stable */
    } else {
        G = KK − MS(WT, 5%);
        foreach v ∈ V_WT do {
            v_decaying stiffness = m';
```

```
            }
        }
}

/∗ step 4 ∗/
clear V_{WT} and E_{WT} in WT;
foreach v ∈ V do {
        add v into V_{WT};
        v_{stiffness} = K / d²_{i,j};
}
compute r for the WT;
tt = 10;
while r ≥ ε do {
        compute r for the WT every tt iteration;
        G = KK − MS(WT, 5%);
}
```
Algorithm 6 Pseudo code of Kamada-Kawai with multi-node selection and decaying stiffness

## 5. EXPERIMENT

To evaluate the proposed algorithms, we implemented a simulation environment using the Java programming language. In a mobile ad hoc network, a number of parameters can influence experimental results, including the total number of nodes in the network, the number of edges per node and the edge distribution within the network. The total number of nodes can affect the performance of force-directed algorithms. In general, the higher the number of nodes in a network topology, the larger the amount of memory consumed and the longer the execution time. Furthermore, the number of edges per node can affect the results. Nodes that have too many edges can affect the performance of the algorithm. Moreover, the variation in the distribution of edges or the non-uniformity of the network topology can affect the results. We conducted our experiments on a computer using an Intel Pentium T2390 processor with 4 GB of memory.

**Table 1 Attributes used in the experiments**

| Type | Parameter | Description | Range |
|---|---|---|---|
| Node distribution | $n$ | Number of nodes in the network | $[2, \infty]$ |
| Edge distribution | $d$ | Average degree of the network | $[2, \infty]$ |
| | $e$ | Edge distribution ratio; lower ratios resulted in sparse networks and higher ratios resulted in uniform networks | $[0.0, 1.0]$ |
| | $\gamma_b$ | Edge generation between a pair of nodes was based on the probability | $[0.0, 1.0]$ |

**Table 2 Performance evaluation criteria**

| Criteria | Description | |
|---|---|---|
| Sensitivity/True positive rate | Boundary nodes on the initial network topology correctly | The higher the better |

| | identified as boundary nodes by algorithms | |
|---|---|---|
| Specificity/True negative rate | Non-boundary nodes on the initial network topology correctly identified as non-boundary nodes by algorithms | The higher the better |
| Execution time | Total amount of execution time that the algorithm ran in seconds | The lower the better |

We considered various network topologies in evaluating the performance and execution time for locating boundary nodes. We adopted the following strategy for network topology generation. First, we defined the attributes for the experiment settings as described in Table 1. This included parameters such as the distributions of nodes and edges. Next, we generated network topologies for the experiments.

For our experiments, we set the total number of nodes ($n$) to 500, 1,000, 2,000, 3,000, 5,000, 7,000 and 10,000. We also set the average degrees of the network ($d$) to 6, 8, 10, 12 and 15 for the generation of network topologies. Overall, 35 network topologies were generated for the experiments based on all of the possible combinations of $n$ and $d$. In addition, all of the network topologies were generated with a node distribution rate ($e$) to 0.25 and edge connectivity ($\gamma_b$) to 0.7.

For our experiments, we focused on network topologies with sparse distributions. Figure 7(a) and (b) illustrate examples of a sparse network and a uniform network, respectively, with the same total number of nodes ($n$) and average degree ($d$). We focused on the network topologies in which nodes were sparsely distributed, as they were more relevant to ad hoc emergency networks.

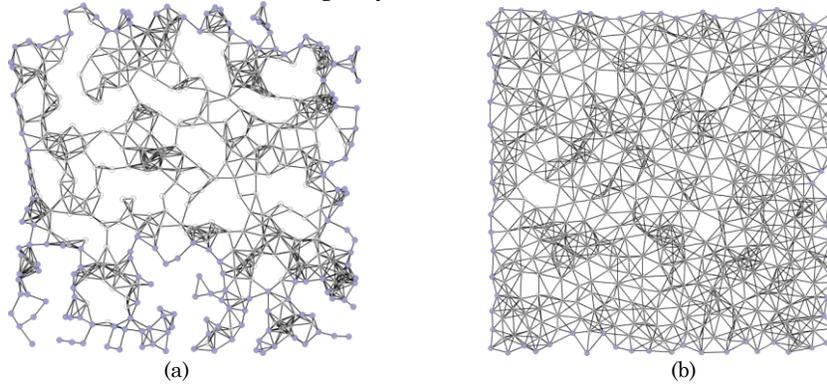

(a) (b)
Figure 7(a) Example of sparse network topology and (b) Example of uniform network topology.

Note that force-directed algorithms are capable of drawing a topology without additional information. According to our evaluation results from Section 3.2, the KK algorithm was able to achieve positive results on topologies with small numbers of nodes. In this experiment, we evaluated the variants of the KK algorithm as proposed in previous sections. These algorithm variants included the original KK algorithm in addition to the KK-SS, KK-MS, and KK-MS-DS algorithms.

In our experiments, we set $K$ to 1. The default size of the drawing frame (canvas) was 600×600. Moreover, the algorithm was allowed to execute for up to 1 minute or until the value of the energy function ($E$) was less than 1.0. According to our extensive testing of the KK algorithm, it was hard to achieve an energy function ($E$) equal to 0. Therefore, we executed each algorithm for 1 minute on all of the network topologies that we generated. For each input network topology, every node was assigned a random visual position within the drawing frame. This random assignment process was repeated five times for each of the thirty-five network topologies to generate one hundred and seventy-five different network topologies.

### 5.1 The Original Kamada-Kawai (KK) Algorithm

We implemented the KK algorithm based on [13]. The corresponding pseudo code is given in Algorithm 2.

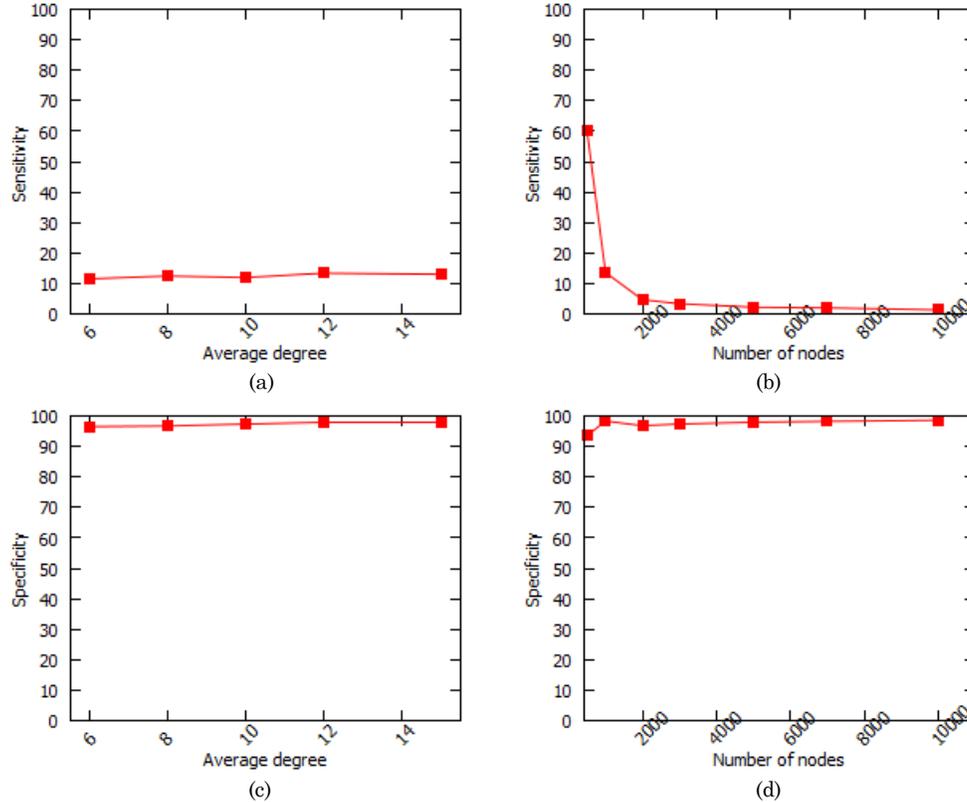

Figure 8(a) Average degree vs. sensitivity, (b) Node count vs. sensitivity, (c) Average degree vs. specificity and (d) Node count vs. specificity.

Figure 8 illustrates the experimental results for the KK algorithm. We calculated the sensitivity and specificity values by averaging the five test samples for each of the thirty-five initial network topologies. They were plotted against the average degree and number of nodes in the network. The sensitivity was stable (approx. 10%) regardless of the average degree. The sensitivity decreased when the number of nodes increased. The experimental results also revealed that specificity was stable regardless of the average degree and number of nodes in the topology.

We did not measure the accuracy of our algorithms. When a large network is used for detection, the number of non-boundary nodes can significantly influence the accuracy. In a network with large numbers of nodes, the number of boundary nodes is much smaller than the number of inner nodes. This means that accuracy depends on the sum of the true positive count (i.e., boundary nodes correctly identified as boundary nodes) and the true negative count (i.e., inner nodes correctly identified as inner nodes) divided by the total number of nodes examined. If the true negative count is high, then the accuracy is high. However, it cannot reflect the real performance of algorithms as applied to boundary detection.

The specificity for networks with fewer than 1,000 nodes was about 4% lower than the other networks. When the number of nodes was less than 1,000, the results differed from those of the rest of the experiment. During the 1-minute algorithm execution time, the results were fairly good with fewer than 1,000 nodes due to the smaller size of the network topologies, although the energy values were greater than 0. However, for the networks with more than 1,000 nodes, the movement of the nodes in the network topologies remained in progress. Here, the shape of the layout was still in its primitive form.

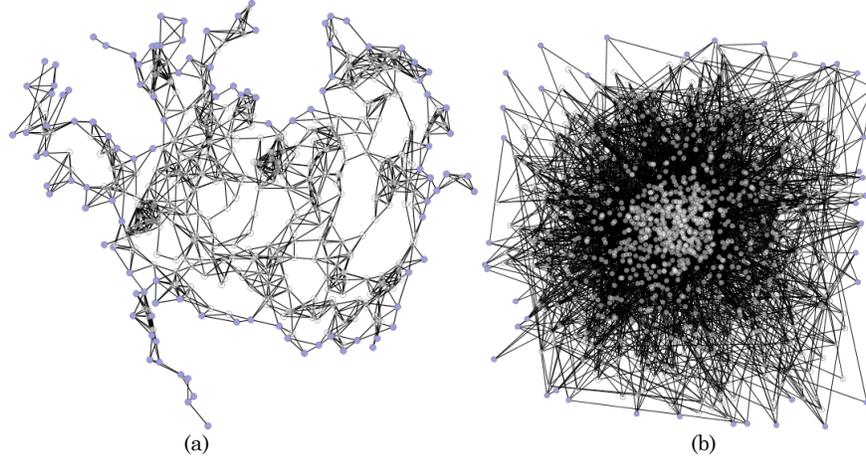

(a) (b)
Figure 9(a) Node count=500 and average degree=6 and (b) Node count=1,000 and average degree=6.

Figure 9 illustrates the case of sufficient expansion based on the two networks with the same average degree. Figure 9(a) is an intermediate result of a test network topology in which the node count is 500 and the average degree is 6. Figure 9(b) is an intermediate result of another test network topology that has the same iteration and average degree as Figure 9(a) but a different number of nodes (node count = 1,000). It is clear that a coarse layout has been forming in Figure 9(a). However, most of the nodes in Figure 9(b) are stacked together. This is caused by the insufficient iteration in Figure 9(b).

### 5.2 Kamada-Kawai with multi-node selection (KK-MS)

We implemented the KK-MS algorithm outlined in Section 4.2. In this experiment, we used the variable $k$ to denote the node percentages, which we selected because they had the highest individual energy value among all of the nodes.

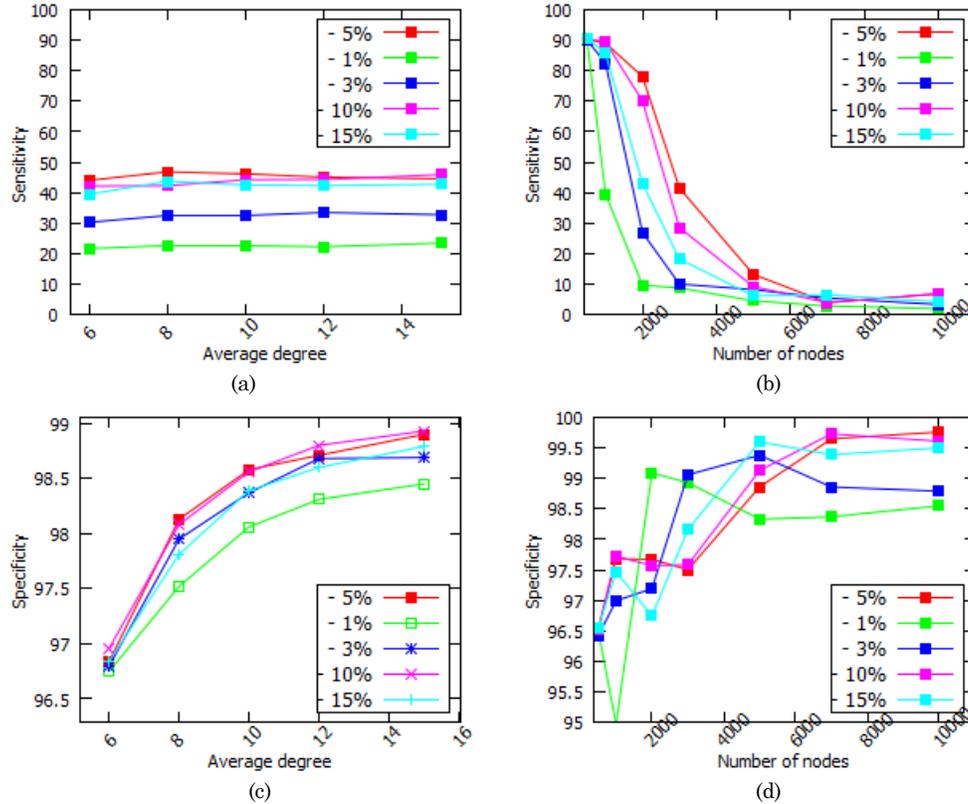

Figure 10(a) Average degree vs. sensitivity, (b) Node count vs. sensitivity, (c) Average degree vs. specificity and (d) Nodes count vs. specificity.

Figure 10 illustrates the experimental results for the KK-MS algorithm. We calculated the sensitivity and specificity values by averaging the five test samples from each of the thirty-five initial network topologies. These are plotted against the average degree and number of nodes in the network.

In the case of the KK-MS algorithm, the level of sensitivity was better than that in the original KK algorithm for all of the test cases when 1%, 3%, 5%, 10% and 15% of the nodes were selected. The specificity was lowest when 1% of the nodes were selected. The drop in specificity was caused by the same problem indicated in Figure 9. A higher percentage of node selection did not necessarily produce good results when the experiments were conducted within a constant execution time. When a large number of nodes were selected, the KK-MS algorithm took on a considerably longer execution time and slowed down the frequency and iteration of the node update. According to our experimental results, the sensitivity associated with selecting 10% or 15% of the nodes was lower than that associated with selecting 5% of the nodes.

Furthermore, our experimental results indicated that when the node count was less than 2,000, selecting 5%, 10% and 15% of the nodes showed a similar sensitivity. Selecting 5% of the nodes produced the best results for the network topologies with node counts between 2,000 and 4,000. However, when the node count was larger than 6,000, the sensitivity results for selecting 5%, 10% and 15% of the nodes were similar.

### 5.3 Kamada-Kawai with signal strength (KK-SS)

Figure 11 shows the experimental results for the KK-SS. The experiment was conducted using the mean of five test samples on thirty-five initial network topologies. The settings for the experiment were the same except for the diameter matrix, which was replaced by signal-strength-based distance. The results indicated that sensitivity and specificity were not affected by signal strength if only the distance measurement of the KK algorithm was replaced.

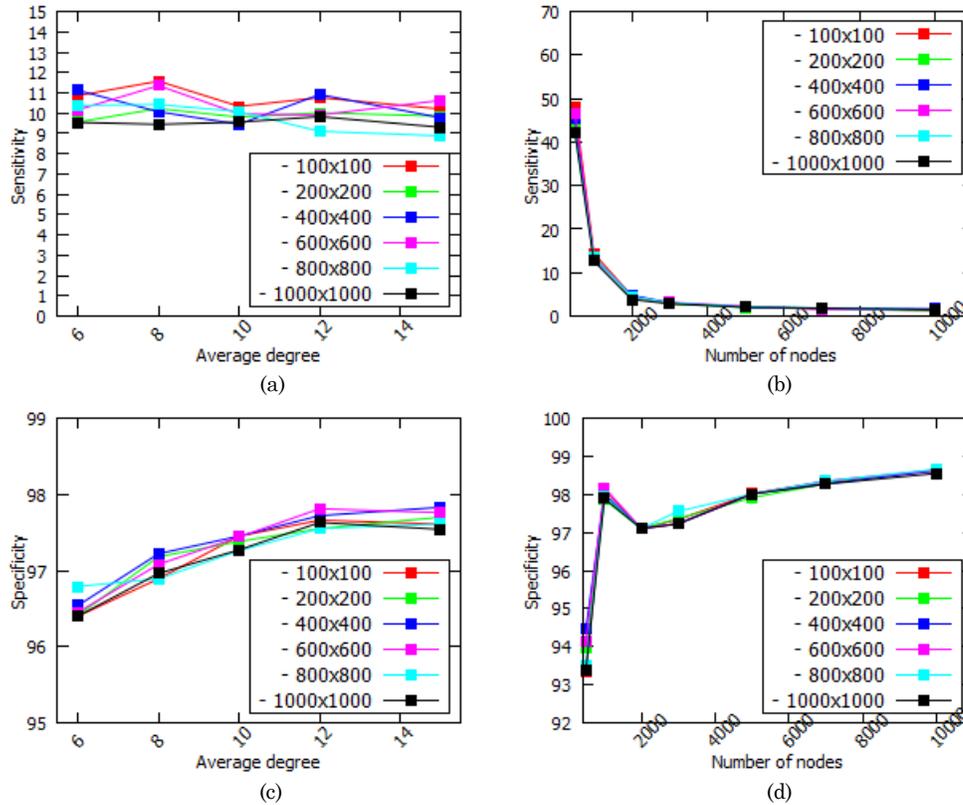

Figure 11(a) Average degree vs. sensitivity, (b) Node count vs. sensitivity, (c) Average degree vs. specificity and (d) Node count vs. specificity.

### 5.4 Kamada-Kawai with multi-node selection and decaying stiffness (KK-MS-DS)

Figure 12 illustrates the experimental results for the KK-MS-DS algorithm, including the sensitivity and specificity results. The experiment was conducted using the mean of five test samples on thirty-five initial network topologies. The KK-MS-DS algorithm achieved the highest level of sensitivity out of all of the KK algorithm variants proposed by this article. The sensitivity result was also stable for all of the choices of average degree.

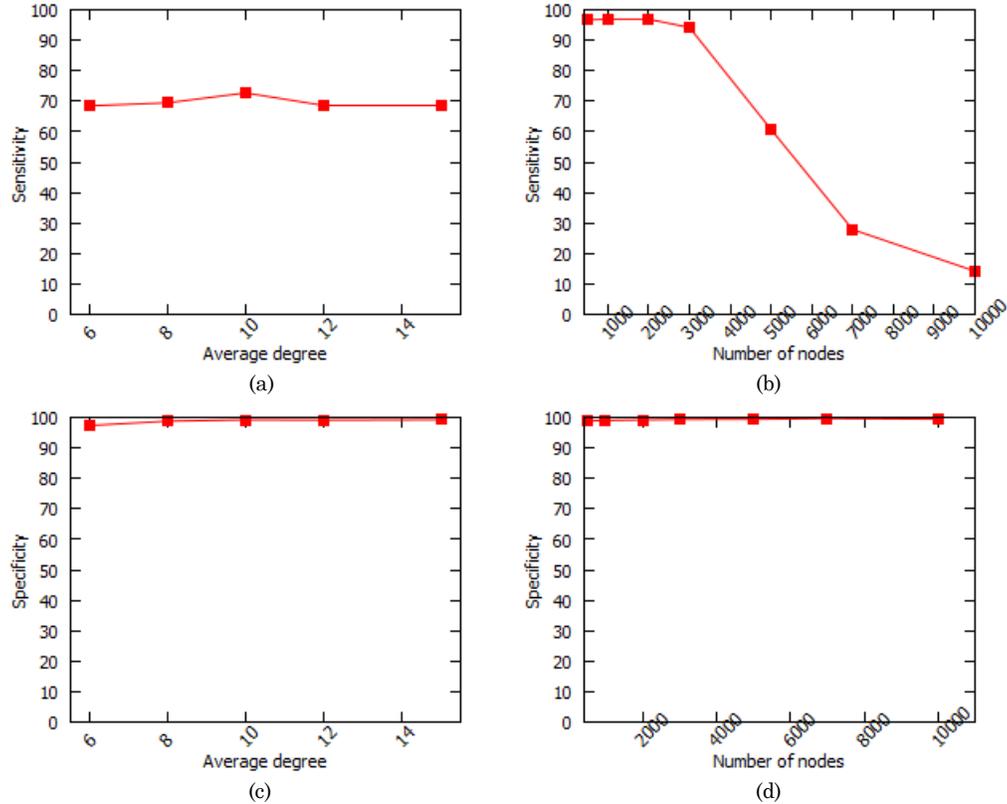

Figure 12(a) Average degree vs. sensitivity, (b) Node count vs. sensitivity, (c) Average degree vs. specificity and (d) Node count vs. specificity.

Furthermore, the KK-MS-DS algorithm achieved 90% sensitivity when the node count was less than 3,000. However, when the node count was larger than 5,000, the sensitivity decreased because the execution time in our experiment was insufficient. Similar to the case described in Section 5.1, most of the nodes remained stacked together. The proposed KK-MS-DS algorithm achieved very high specificity (almost 100%) in all of the cases in the experiment models.

### 5.5 Overall comparison

Figure 13 illustrates the overall results of the KK, KK-MS and KK-MS-DS algorithms for determining sensitivity and specificity. From these results, we can observe that the KK-MS-DS algorithm had the highest sensitivity and specificity out of all of the approaches. The KK-MS-DS algorithm also had higher and more stable results for specificity than the other approaches. The specificity of the KK and KK-MS algorithms dropped when the node count reached 2,000.

Accelerating Kamada-Kawai for boundary detection in mobile ad hoc network 39:26

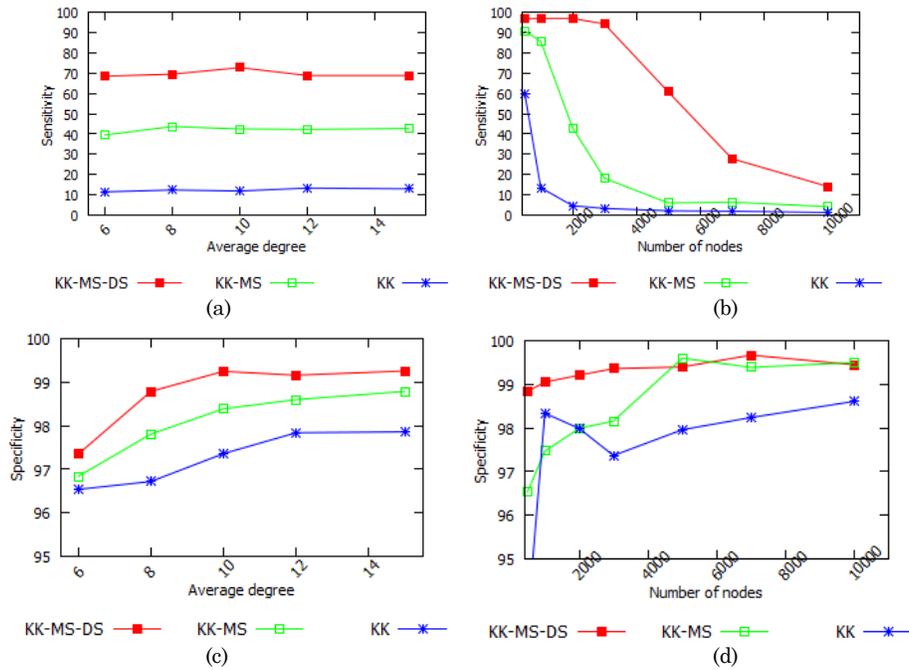

Figure 13(a) Average degree vs. sensitivity, (b) Node count vs. sensitivity, (c) Average degree vs. specificity and (d) Node count vs. specificity.

### 5.6 Comparison based on a test case

In this section, we compare the proposed approaches and analyse the advantages and disadvantages associated with each approach using a particular test case with a node count $n = 500$. Figure 14 illustrates the experimental results on networks with 500 nodes and average degrees of 6, 8, 10, 12 and 15. The execution time was set to 4 seconds.

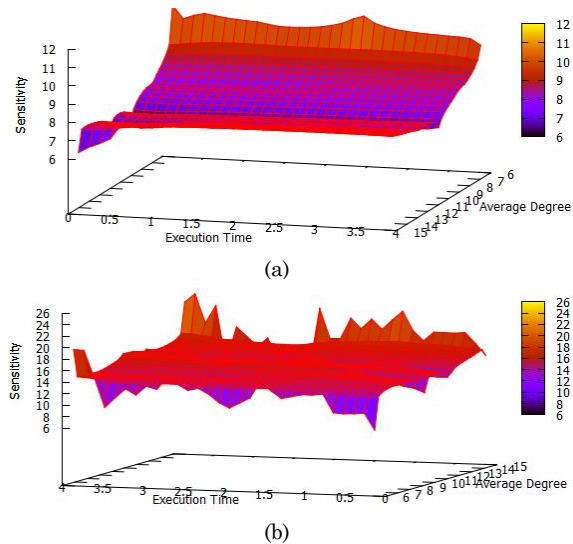

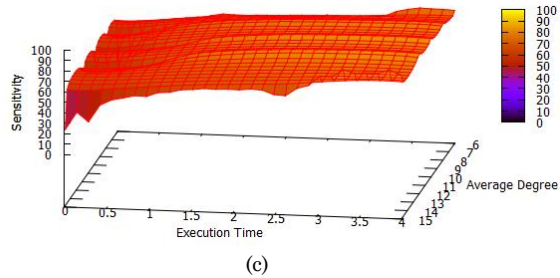

(c)

Figure 14 Sensitivity vs. Average degree vs. Execution time trend for the KK algorithm with a node count=500, (b) Sensitivity vs. Average degree vs. Execution time trend for the KK-MS algorithm with a node count=500 and (c) Sensitivity vs. Average degree vs. Execution time trend for the KK-MS-DS algorithm with a node count=500.

According to Figure 14(a), the original KK algorithm achieved a better performance when the average degree was lower. The highest degree of sensitivity it obtained was 12%. In contrast to the original KK algorithm, the results for the KK-MS algorithm with $k = 5$ from Figure 14(b) indicate that the degree of sensitivity was low when the average degree of the network was low. However, when the average degree was higher, the sensitivity increased up to 26%. The results for the KK-MS-DS algorithm with $k = 5$ from Figure 14(c) achieved the highest and most relatively stable sensitivity out of all of the algorithms.

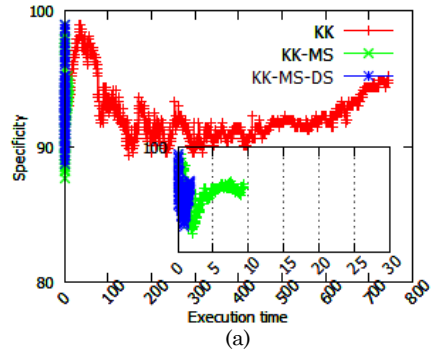

(a)

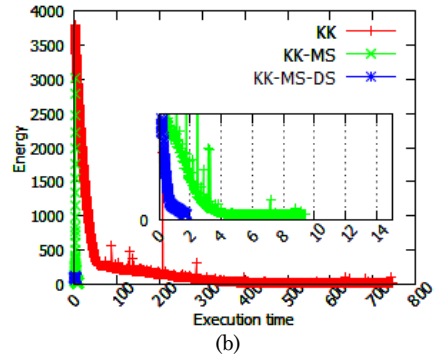

(b)

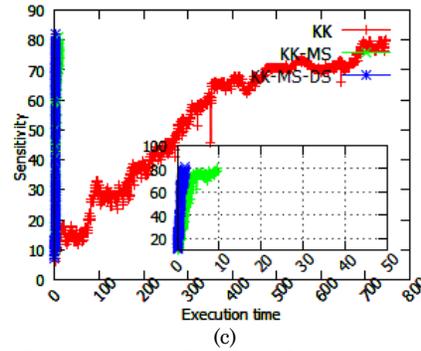
(c)

Figure 15(a) Specificity with node count=500 and average degree=8, (b) Energy with node count=500 and average degree=8 and (c) Sensitivity with node count=500 and average degree=8.

Figure 15(b) shows the average value of the energy function E for the experiments. Some of the key parts of these charts are magnified for a better illustration. According to our experimental results, the KK-MS-DS algorithm achieved the best results. It resulted in the fastest reduction in the energy function value ($E$). In KK, it took 745.40 time units to reduce the the energy value to 23.29. In KK-MS-DS, it took just 1.78 time unit to reach the energy value at 14.84. KK-MS-DS is approximately 418 times faster than original KK algorithm. Therefore, the experimental results indicate that the KK-MS and KK-MS-DS algorithms showed a faster convergence on the energy function ($E$) than the KK algorithm. Furthermore, according to the principle of the KK algorithm, the smaller the value of the energy function ($E$), the better the layout (i.e., less edge crossing and more planar in shape).

According to Figure 15(a) and (c), the KK-MS-DS algorithm also achieved a much faster convergence rate in terms of sensitivity and specificity. In original KK, it took 745.40 time units to achieve 80% sensitivity and specificity where as in KK-MS-DS, it just took 1.78 time units for the same measures. In this case, KK-MS-DS is also 418 times faster than original KK.

## 6. CONCLUSION AND FUTURE WORK

We propose a novel approach to applying the KK algorithm to boundary node detection in mobile ad hoc networks. Based on our extensive testing of three classical force-directed algorithms, we select the KK algorithm to further improve its accuracy and convergence rate. First, we propose an algorithm known as the KK-MS algorithm to perform multi-node selection for the KK algorithm and thereby speed up its updating procedure on the network. We also analyse the relationship between signal strength, edge length and convergence for the KK algorithm. Based on this analysis, we propose a heuristics KK-MS-DS algorithm that estimates distance estimation by signal strength, replacing the edge length from the hop count to the estimated distance. Moreover, we use decaying stiffness to fine-tune the force acting on every node within the network topologies. The KK-MS-DS algorithm provides a dynamic update for decaying stiffness to achieve rapid convergence. The algorithm encourages the effective 'expansion' of would-be boundary nodes in an outward direction. Our experimental results show that the improved algorithm can significantly shorten the processing time and detect boundary nodes with an acceptable level of accuracy. Testing on a 500 nodes network shows that KK-MS-DS is approximately 400 times faster than original KK in reducing energy as well as in increasing sensitivity and specificity measures.

In our future work, we plan to implement a distributed version of the KK-MS-DS algorithm for mobile ad hoc networks. This distributed KK-MS-DS algorithm could be used in situations where the nodes within a network have limited processing power and only topological information is available. For instance, each node (censor) within a mobile ad hoc network may only be aware of the existence of its nearest neighbours (e.g., two-hops neighbours) and centralised processing may be impossible due to hardware limitations. In this case, a distributed KK-MS-DS algorithm could be used to determine the boundary nodes of the network. One of the main objectives in distinguishing the boundary nodes from the rest is to prolong their battery life as long as possible. Boundary nodes are crucial for locating the existence of a network and for maintaining communication with other networks. The emergency depicted in Lifeline comprises a scenario highlighting the use of mobile ad hoc networks [1].

ACKNOWLEDGEMENT

This research was funded by the Research Committee of University of Macau, grant MYRG041(Y1-L1)-FST13-SYW and MYRG2015-00054-FST.